\documentclass[11pt,preprint]{aastex}
\usepackage{multicol}

\slugcomment{ }

\shorttitle{ }
\shortauthors{ }
\newcommand{\fpr}[1]{\left( #1\right)}

\newcommand{\fang}[1]{\langle #1\rangle}
\newcommand{\fabs}[1]{\left| #1\right|}
\newcommand{\muK}{{\mu\mathrm{K}}}
\begin{document}
\title{The Validity of the Cosmic String Pattern Search with the Cosmic 
Microwave Background \\}
\author{E. Jeong}
\affil{Department of Physics, University of California, Berkeley, CA, 94720}
\email{ehjeong@berkeley.edu}
\author{G. F. Smoot}
\affil{Department of Physics, University of California, Berkeley, CA, 94720}
\affil{Lawrence Berkeley National Laboratory, Berkeley, CA 94720}
\email{gfsmoot@lbl.gov}
\begin{abstract}
We introduce a new technique to detect the discrete temperature steps that 
cosmic strings might have left in the cosmic microwave background (CMB) 
anisotropy map. The technique provides a validity test on the pattern search of
cosmic strings that could serve as the groundwork for future pattern searches. 
The detecting power of the technique is only constrained by two unavoidable 
features of CMB data: (1) the finite pixelization of the sky map and (2) the 
Gaussian fluctuation from instrumental noise and primordial anisotropy. We set 
the upper limit on the cosmic string parameter as 
$G\mu\lesssim 3.7\times 10^{-6}$ at the 95\% confidence level (CL) and find 
that the amplitude of the temperature step has to be greater than $44\mu K$ in 
order to be detected for the {\it{Wilkinson Microwave Anisotropy Probe (WMAP)}}
3 year data.
\end{abstract}


\section{Introduction and Modeling}
Cosmic string is one of the relic structures that are predicted to be produced 
in the course of symmetry breaking in the hot, early universe, whose discovery 
will probably be an important landmark for the high energy physics. The quest 
for cosmic strings has been conducted in two ways: statistical method and 
direct search for individual cosmic string. On the statistical side, many 
studies agreed that the contribution from cosmic strings to statistical 
observables such as the power spectrum is at most $10\%$ 
\citep{pogosian.et.al,wu,fraisse,daviskibble,wyman.et.al}, reconfirming that 
cosmic strings played a minor role, if any, in making the universe. Other 
workers have set upper limits on the cosmic string parameter $G\mu$ 
\citep{hindmarsh1, perivolaropoulossimatos,jeongsmoot1,lowright,seljakslosar,
pogosian.et.al2,fraisse2}. In this Letter, we introduce a technique for 
estimating how strong a signal from cosmic strings has to be in order to be 
identified unequivocally. 
The detecting power of this technique is calibrated by applying it to 
simulated anisotropy maps based on reasonable modeling. In the last part of 
this Letter, we apply this technique to the {\it{WMAP}} 3 year W-band data 
set\footnote{See http://lambda.gsfc.nasa.gov/product/map/} and analyze the
implication of the results.
A cosmic string can leave discrete temperature steps in a CMB anisotropy map 
due to the Kaiser-Stebbins effect \citep{kaiserstebbins}, with the height of a 
step $\delta T$ given by
\begin{equation}\label{1}
\delta T= 8\pi G\mu\gamma_s\beta_sT\hat{n}\cdot (\hat{v}\times\hat{s})
\end{equation}
where $T\simeq 2.725 K$ is the universal background temperature of CMB. 
However, those step patterns are probably obscured by instrumental noises and 
other physical structures of anisotropies. 
\begin{figure}[t]
\begin{center}
\includegraphics[width=13cm,angle=0]{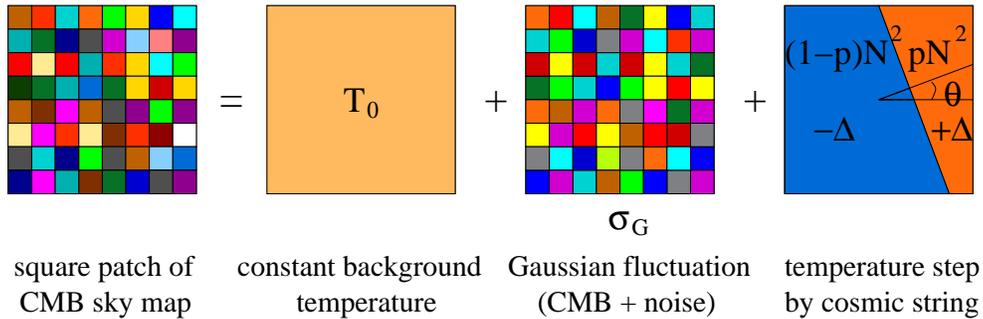}
\caption{\label{fig1}Decomposition of a CMB anisotropy sky map patch. A 
square patch is decomposed into the uniform background temperature ($T_0$) + 
Gaussian fluctuation ($\sigma_G$)+ discrete temperature step ($\pm\Delta$). 
The number of pixels in a patch is $N^2$.}
\end{center}
\end{figure}
We consider a square patch of the CMB anisotropy sky map that extends to a 
size of the horizon at the time of recombination so that the conic spacetime 
formed by a cosmic string can be in effect within the region. Then another 
square patch of the CMB anisotropy sky map that contains a segment of moving 
cosmic string can be decomposed into three parts, as illustrated in Figure 
\ref{fig1}: (1) the uniform background temperature ($T_0$), (2) the Gaussian 
fluctuation (variance $\sigma_G^2$), and (3) the discrete temperature step 
($\pm\Delta$) from a moving cosmic string. The uniform background temperature 
$T_0$ comes from superhorizon-scale primordial fluctuations. Since we pick a 
horizon-sized region, the super-horizon fluctuations will appear to be a 
constant temperature shift for the whole patch. Smaller scale (sub-horizon) 
fluctuations and instrumental noise add up incoherently to form a Gaussian 
fluctuation with the variance $\sigma_G^2=\sigma_{CMB}^2+\sigma_{noise}^2$, 
where $\sigma_{CMB}^2$ and $\sigma_{noise}^2$ are variances for fluctuations 
of primordial origin and instrumental noise, respectively. We introduce five 
parameters that characterize a square patch of a string-embedded sky map, 
$T_0$, $\sigma_G$, $\Delta$, $p$ (blueshifted pixels/total pixels), and 
$\theta$ (orientation of step). We concoct a simulated patch of the CMB 
anisotropy map by 
assigning arbitrary legitimate values to the parameters and adding the three 
components illustrated in Figure \ref{fig1}. To recover these parameters from 
the CMB anisotropy map where the step pattern is intermixed, we employ five 
observables $(p,\:\Delta,\:\theta,\:\sigma_G,\:T_0)$ that can be expressed in 
terms of 
\begin{eqnarray}
{\mathrm{mean}}&:&\tau=N^{-2}\sum_iT_i\label{2}\\
{\mathrm{dipole\: moment}}&:&{\bf{d}}=N^{-3}\sum_iT_i{\bf{r}}_i\label{3}\\
{\lambda -\mathrm{inertia}}&:&\lambda =4N^{-3}\sum_i
T_i\fabs{{\bf{d\cdot r}}_i}/\fabs{{\bf{d}}}\label{4}\\
{\mathrm{variance}}&:&\sigma^2=N^{-2}\sum_i\fpr{T_i-\tau}^2\label{5}
\end{eqnarray}
where $T_i$ represents the signal at the $i$th pixel and ${\bf{r}}_i$ 
is the two-dimensional gridded position vector for the $i$th pixel 
defined in a square patch. 
\begin{figure}[t]
\begin{center}
\includegraphics[width=13cm,angle=0]{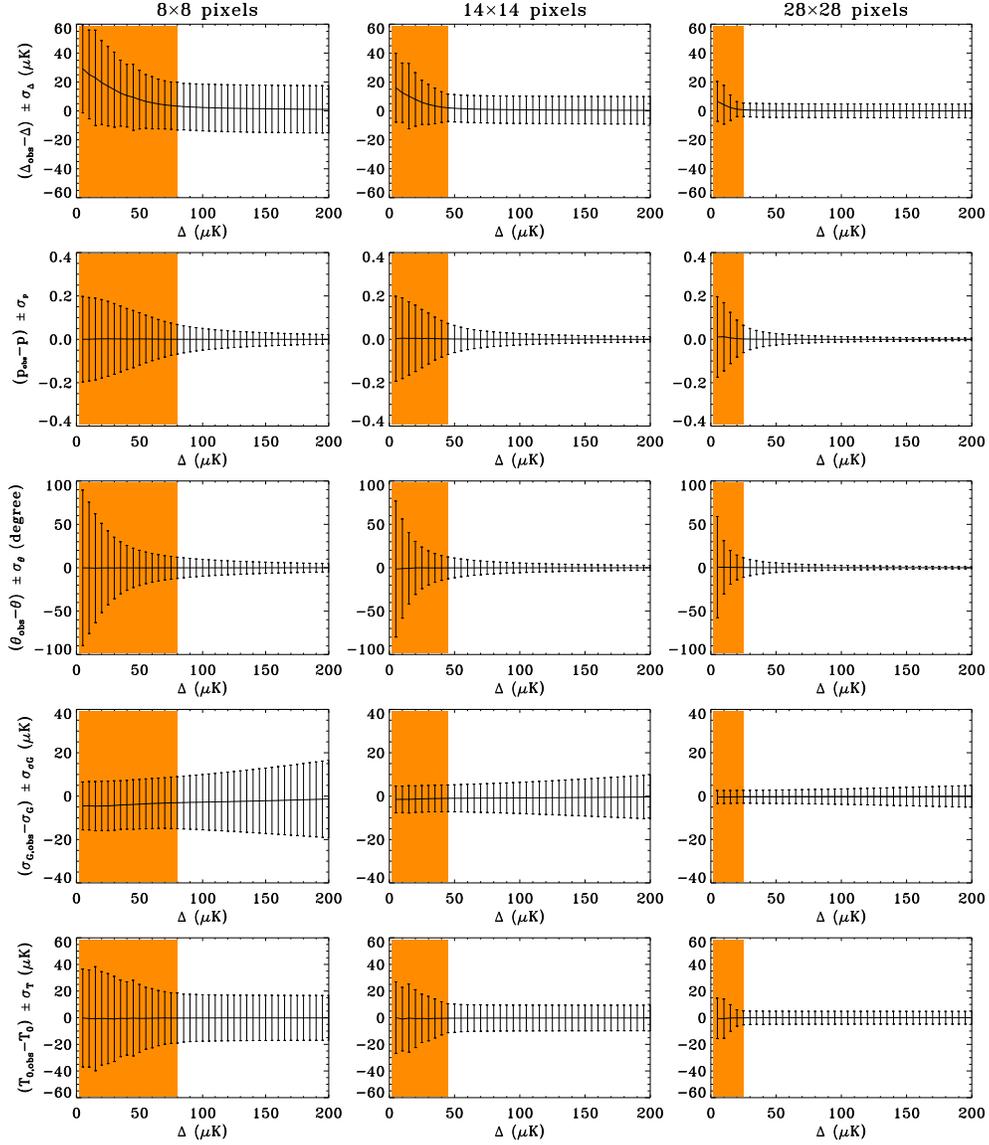}
\caption{\label{fig2}Evolution of errors of characteristic parameters as 
functions of $\Delta$ with input values of parameters 
$\fpr{T_0=10.0\mu K,\: p=0.5,\:\theta =0,\:\sigma_G=116.0\mu K}$.
{\it{Left panels}}: $8\times 8$ pixels in a square. {\it{Middle panels}}: 
$14\times 14$ pixels. {\it{Right panels}}: $28\times 28$ pixels. For each size 
of patch, the number of simulations performed is 10,000. The shaded regions 
indicate the ranges of $\Delta$ for which the step feature in a patch is not 
obvious compared to its background fluctuation ($\sigma_G=116.0\;\muK$). The 
result on $\Delta_{obs}$ is biased upward because the simulations with 
erroneous values of $\Delta_{obs}$ ($\Delta_{obs} < 0$) were not included in 
the statistics.}
\end{center}
\end{figure}
In the absence of the background fluctuation ($\sigma_G=0$), the following 
analytic expressions for the characteristic parameters return the exact input 
values:
\begin{eqnarray}
p&=&\frac{2d -\frac{1}{2}\fpr{\fabs{\lambda-\tau}+\lambda-\tau}}
{4d -\fabs{\lambda-\tau}}\label{6}\\
\Delta &=&\frac{\fpr{\fabs{\lambda-\tau}-4d}^2}
{2\fpr{2d-\fabs{\lambda-\tau}}}\label{7}\\
\theta &=&\tan^{-1}\fpr{d_y/d_x}\label{8}\\
\sigma_G&=&\fpr{\sigma^2-4\Delta^2p\fpr{1-p}}^{1/2}\label{9}\\
T_0 &=&\tau-\Delta\fpr{2p-1}.\label{10}
\end{eqnarray}
With the background fluctuation turned on, an observed characteristic parameter 
${\mathcal{Q}}_{i,obs}$ is expressed as an unbiased estimate 
${\mathcal{Q}}_{i}$ plus a Gaussian error $\sigma_{{\mathcal{Q}}_{i}}$ 
$\fpr{{\mathcal{Q}}_{i}=p,\:\Delta,\:\theta,\:\sigma_G,\:T_0}$,
\begin{equation}\label{11}
{\mathcal{Q}}_{i,obs}={\mathcal{Q}}_{i}\pm \sigma_{{\mathcal{Q}}_{i}}.
\end{equation}
The errors ($\sigma_{p},\sigma_{\Delta},\sigma_{\theta},\sigma_{\sigma_G},
\sigma_{T_0}$) present in the equations (\ref{11}) are measures of how 
precisely the information on a temperature step screened by background 
fluctuation is recovered. Figure \ref{fig2} shows the behaviors of errors as 
functions of $\Delta$ for three different pixel numbers for a patch. The 
standard deviation of the background fluctuation chosen here is the mean 
standard deviation of patches with angular radius $\theta_R=1.8^{\circ}$ (the 
angular size of the horizon at the recombination for the $\Lambda$CDM model) 
for the WMAP 3 year W-band data set with the Kp2 mask applied. The most 
pronounced feature of the graphs is that the error bars are wildly undulating 
for small $\Delta$ compared to $\sigma_G$, with attenuating envelopes as 
$\Delta$ increases. The algorithm works poorly for low $\Delta$ compared to 
$\sigma_G$, such as the $\pm 90^{\circ}$ error on the orientation estimate.
\begin{table}[th]
\begin{center}
\begin{tabular}{cr@{.}lr@{.}l@{ $\pm$ }r@{.}l}
\hline\hline
\hspace*{1cm}Parameter\hspace*{1cm}&\multicolumn{2}{c}{Input}&\multicolumn{4}{c}{\hspace*{1cm}Output (1 $\sigma$)\hspace*{1cm}}\\
\hline
  &\multicolumn{2}{c}{Patch with $8\times 8$ pixels}&\multicolumn{4}{c}{ }\\
\hline
$p$\dotfill&\hspace*{1.5cm}0&5&\hspace*{1cm}0&5&0&1\\
$\Delta\:\fpr{\muK}$\dotfill&\hspace*{1.5cm}80&0&\hspace*{1cm}83&4&16&6\\
$\theta$ (deg)\dotfill&\hspace*{1.5cm}0&0&\hspace*{1cm}-0&1&12&3\\
$\sigma_G (\muK )$\dotfill&\hspace*{1.5cm}116&0&\hspace*{1cm}113&0&12&0\\
$T_0 (\muK )$\dotfill&\hspace*{1.5cm}10&0&\hspace*{1cm}9&7&18&9\\
\hline
  &\multicolumn{2}{c}{Patch with $14\times 14$ pixels}&\multicolumn{4}{c}{ }\\
\hline
$p$\dotfill&\hspace*{1.5cm}0&5&\hspace*{1cm}0&5&0&1\\
$\Delta\:\fpr{\muK}$\dotfill&\hspace*{1.5cm}45&0&\hspace*{1cm}47&1&9&5\\
$\theta$ (deg)\dotfill&\hspace*{1.5cm}0&0&\hspace*{1cm}-0&2&12&6\\
$\sigma_G (\muK )$\dotfill&\hspace*{1.5cm}116&0&\hspace*{1cm}115&0&6&2\\
$T_0 (\muK )$\dotfill&\hspace*{1.5cm}10&0&\hspace*{1cm}9&6&10&9\\
\hline
  &\multicolumn{2}{c}{Patch with $28\times 28$ pixels}&\multicolumn{4}{c}{ }\\
\hline
$p$\dotfill&\hspace*{1.5cm}0&5&\hspace*{1cm}0&5&0&1\\
$\Delta\:\fpr{\muK}$\dotfill&\hspace*{1.5cm}25&0&\hspace*{1cm}25&8&4&7\\
$\theta$ (deg)\dotfill&\hspace*{1.5cm}0&0&\hspace*{1cm}0&2&11&2\\
$\sigma_G (\muK )$\dotfill&\hspace*{1.5cm}116&0&\hspace*{1cm}115&7&3&0\\
$T_0 (\muK )$\dotfill&\hspace*{1.5cm}10&0&\hspace*{1cm}9&9&5&3\\
\hline\hline
\end{tabular}
\end{center}
\caption{\label{table1}Simulation results at the critical values of
$\Delta$ above which $\sigma_{\Delta}$s are small enough and do not decrease
any more.}
\end{table}
Even the breakdowns of the algorithm do happen in the shaded region, 
resulting in erroneous parameter values such as negative $\Delta$ or $p$ not 
in the range between 0 and 1. Simulations with collapsed results are not 
included in the statistics shown in Figure \ref{fig2} since those cases are 
evidently the ones with insufficiently strong signals against background 
fluctuation. As $\Delta$ increases, we begin obtaining the computed parameters 
that are very close to the true values with acceptable errors (dubbed the 
``good'' results), and it allows us to recover the temperature step parameters 
faithfully. We also find from Figure \ref{fig2} that a patch with more pixels 
works better with a faster attenuation of uncertainties. Table \ref{table1} 
displays the performances of the algorithm for the values of $\Delta$ at the 
borders, above which errors for the temperature step ($\sigma_{\Delta}$) do not
get any better. We use circular patches for {\it{WMAP}} data analysis rather 
than square patches because of the computational advantage. Algebraic 
expressions for the observables given in equations (\ref{6})-(\ref{10}) work 
very well for circular patches with negligible numerical differences for the 
case of square patches. A square patch with $28\times 28$ pixels at the normal 
resolution of {\it{WMAP}} W-band data covers nearly the same area as the 
circular region with angular radius $\theta_R=1.8^{\circ}$. We performed a 
further detecting power test with $28\times 28$ pixels patches and found 
empirically that the relation between the critical value of $\Delta$, 
$\Delta_c$, above which ``good'' results start to come out, and $\sigma_G$ is
\begin{equation}\label{12}
\Delta_c\simeq 0.25\sigma_G.
\end{equation}
The different choices of input values for the parameters $p$, $\theta$, or 
$T_0$ showed no noticeable difference in the evolutions of errors.
\section{Application to {\it{WMAP}}}
An observed CMB anisotropy map is an aggregate of various independent modes of
perturbation ranging from tiny sub-horizon scales to super-horizon scales well
beyond the correlation length. As illustrated in Figure \ref{fig1},
fluctuations with larger or smaller scales compared to the size of a test
patch are neatly prescribed, but the intermediate scales whose wavelengths
are comparable to the size of a test patch will appear to be continuous
temperature tilts that also give plausible values of the step parameters. One
drawback of this algorithm is that it does not distinguish between a discrete
temperature step and a smooth temperature slope. However, this shortcoming can
be easily fixed: If an apparent temperature step is detected at a spot on the 
map, we repeat the analysis with a half-sized patch at the same spot. If the 
structure is a continuous slope, then it would return half the value of 
$\Delta_{obs}$ than the previous result, while for the signal from a discrete 
step, the returned $\Delta_{obs}$ should stay the same within the error. We 
conducted the step signal search through the {\it{WMAP}} 3 year W-band data set
and found 193,160 unqualified signals against background fluctuation (the 
constraint in eq. [\ref{12}] is not applied), 129,049 qualified steps+tilts 
(signals that meet the constraint eq. [\ref{12}]; i.e., 
$\Delta_{obs}\ge 0.25\sigma_G$), and 12,330 qualified discriminated steps. 
\begin{figure}[t]
\begin{center}
\includegraphics[width=7cm,angle=0]{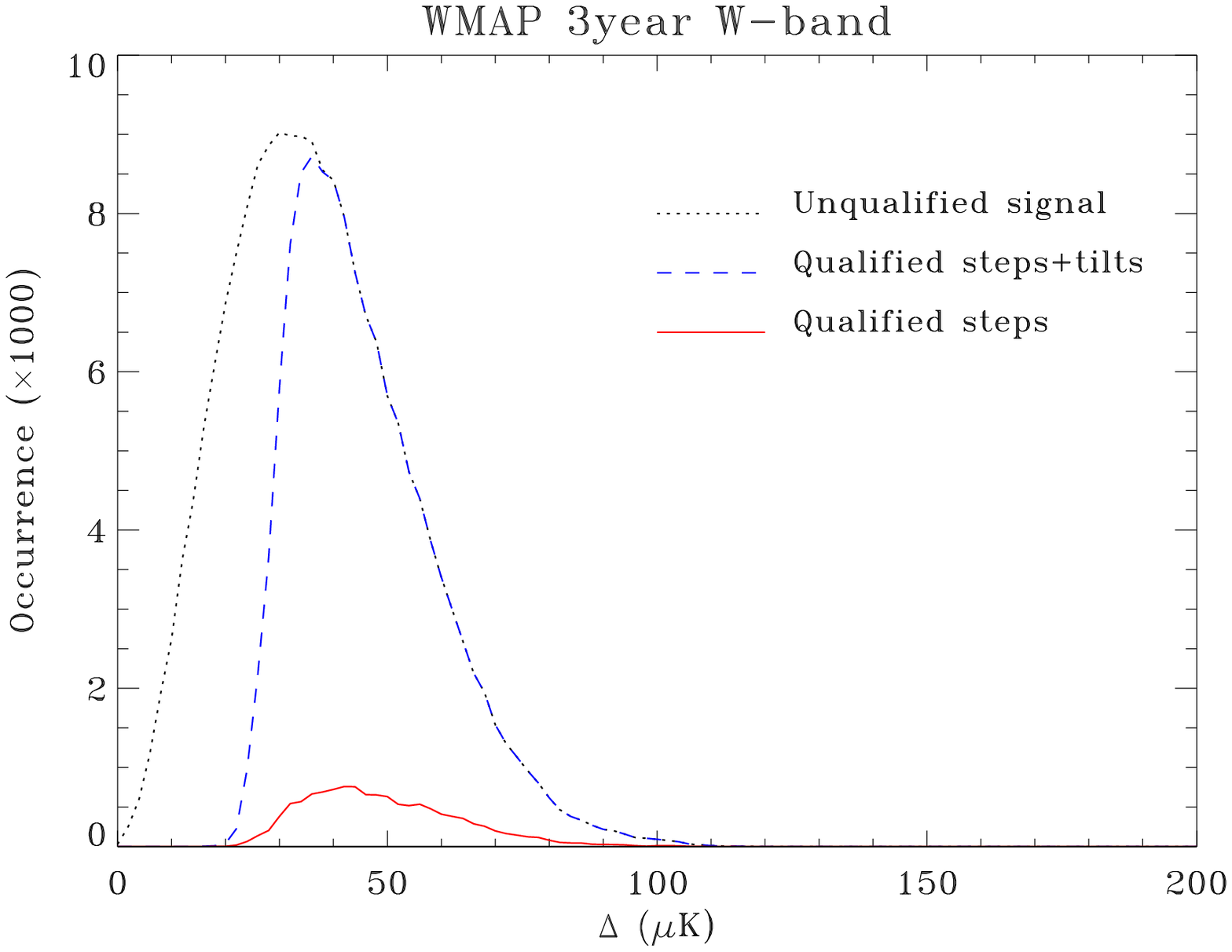}
\end{center}
\caption{\label{fig3}Distributions of $\Delta_{obs}$ detected in {\it{WMAP}} 
3 year W-band data set. The number of step signals found are 193,160 
unqualified signals ({\it{dotted curve}}), 129,049 qualified steps+tilts 
({\it{dashed curve}}), and 12,330 qualified steps ({\it{solid curve}}).}
\end{figure}
The $\Delta_{obs}$ value of qualified steps is in the range
\begin{equation}\label{13}
(18.3\pm 3.0)\mu K< \Delta_{obs} < (115.4\pm 6.0)\mu K
\end{equation}
as the curves show in Figure \ref{fig3}.
Patches with radius $\theta_R=1.8^{\circ}$ in the {\it{WMAP}} 3 year W-band 
have a background fluctuation $\sigma_G$ less than 176 $\muK$ in the 
Kp2-mask-cleared region. This means, at the worst case, we can identify a step 
signal as low as $176\;\muK /4=44\;\muK$ with moderate errors. Therefore, if 
there are cosmic string signals with $\Delta \ge 115.4\;\muK$ and if they are 
located in the available region (out of the Galactic plane or the Kp2 masked 
region), they would not be missed. Thus, we can set an upper limit on the 
cosmic string signal
\begin{equation}\label{14}
\Delta =4\pi G\mu\gamma_s\beta_sT|\cos\phi |<127.4\mu K,\quad 95\%
\:{\mathrm{CL}}
\end{equation}
where $\phi$ is arbitrary. Thus, with $\fang{\gamma_s\beta_s}\simeq 1$
\citep{vilenkinshellard:CS}, the upper limit of the cosmic string parameter 
$G\mu$ can be estimated as
\begin{equation}\label{15}
G\mu\lesssim 3.7\times 10^{-6},\quad 95\%\: {\mathrm{CL}}.
\end{equation}
\section{CONCLUSIONS}
We have presented and tested a new technique to directly discover cosmic 
strings via the patterns they would produce in a CMB anisotropy map. We found 
that the minimum magnitude of the step signal that is required to be 
unequivocally identified is $44\;\muK$ for the {\it{WMAP}} 3 year W-band data 
set. This algorithm can be used to crop the reliable step signals from the CMB 
anisotropy data, and it will serve as the valuable ground work for future
pattern searches with more refined data, such as further {\it{WMAP}} data 
releases or {\it{PLANCK}} data.

Computer simulations and data analysis with the {\it{WMAP}} data set were done 
using HEALPix\footnote{See http://healpix.jpl.nasa.gov} \citep{Gorski.et.al}. 
This work was supported by LBNL and the Department of Physics at the 
University of California, Berkeley.


\begin{thebibliography}{99}
\bibitem[Pogosian et al. 2003]{pogosian.et.al}Pogosian et al., 2003,
Phys.Rev.D, 68, 023506
\bibitem[Wu 2005]{wu}Wu, J-H., 2005, astro-ph/0501239
\bibitem[Fraisse 2005]{fraisse}Fraisse, A., 2005, astro-ph/0503402
\bibitem[Davis \& Kibble 2005]{daviskibble}Davis, A.-C. \& Kibble, T.W.B, 2005, 
Contemp.Phys, 46, 313
\bibitem[Wyman et al. 2005]{wyman.et.al}Wyman, M., Pogosian,L., \& 
Wasserman, I., 2005, Phys.Rev.D, 72, 023513
\bibitem[Hindmarsh 1994]{hindmarsh1}Hindmarsh, M., 1994, ApJ, 431, 534
\bibitem[Perivolaropoulos \& Simatos 1998]{perivolaropoulossimatos}
Perivolaropoulos, L. \& Simatos, N., 1998, astro-ph/9803321
\bibitem[Jeong \& Smoot 2005]{jeongsmoot1}Jeong, E. \& Smoot, G. F., 2005,
ApJ, 624, 21
\bibitem[Lo \& Wright 2005]{lowright}Lo, A. S. \& Wright, E. L., 2005, 
astro-ph/0503120
\bibitem[Seljak \& Slosar 2006]{seljakslosar}Seljak, U. \& Slosar, A., 2006, 
Phys.Rev.D, 74, 063523
\bibitem[Pogosian et al. 2006]{pogosian.et.al2}Pogosian, L., Wasserman, I., 
\& Wyman, M., 2006, astro-ph/0604141
\bibitem[Fraisse 2006]{fraisse2}Fraisse, A., 2006, astro-ph/0603589
\bibitem[Kaiser \& Stebbins 1984]{kaiserstebbins}Kaiser, N. \& Stebbins, A., 
1984, Nature, 310, 391
\bibitem[Vilenkin \& Shellard 1994]{vilenkinshellard:CS}Vilenkin, A. \& 
Shellard, E. P. S., 1994, Cosmic Strings and Other Topological Defects, 
(Cambridge, UK, Cambridge University Press)
\bibitem[G\'{o}rski et al. 2005]{Gorski.et.al}G\'{o}rski, K. M. et.al., 2005,
ApJ, 622, 759
\end{thebibliography}
\end{document}